\documentclass[prl,twocolumn,amssymb,showpacs,floatfix,%
superscriptaddress]{revtex4}  

\usepackage{graphicx}

\begin{document}

\title{Electron capture rates on nuclei and implications for stellar
  core collapse} 
\author{K. Langanke}
\affiliation{Institute for Physics and Astronomy, 
University of {\AA}rhus,
  DK-8000 {\AA}rhus C, Denmark}
\author{G. Mart\'{\i}nez-Pinedo}
\affiliation{Institut d'Estudis Espacials de Catalunya, 
Edifici Nexus, Gran Capit\`a 2, 
E-08034 Barcelona, Spain}
\affiliation{Instituci\'o Catalana de Recerca i Estudis Avan\c{c}ats,
Llu\'{\i}s Companys 23, 
E-08010 Barcelona, Spain}
\author{J. M. Sampaio}
\affiliation{Institute for Physics and Astronomy, 
University of {\AA}rhus,
  DK-8000 {\AA}rhus C, Denmark}
\author{D. J. Dean}
\affiliation{Physics Division, Oak Ridge National Laboratory, Oak
  Ridge, TN 37831}
\author{W.~R.~Hix}
\affiliation{Physics Division, Oak Ridge National Laboratory, Oak
  Ridge, TN 37831}
\affiliation{Department of Physics and Astronomy, University of
  Tennessee, Knoxville TN 37996}
\affiliation{Joint Institute for Heavy Ion research, Oak Ridge, TN
  37831}
\author{O.~E.~B.~Messer}
\affiliation{Physics Division, Oak Ridge National Laboratory, Oak
  Ridge, TN 37831}
\affiliation{Department of Physics and Astronomy, University of
  Tennessee, Knoxville TN 37996}
\affiliation{Joint Institute for Heavy Ion research, Oak Ridge, TN
  37831}
\author{A. Mezzacappa}
\affiliation{Physics Division, Oak Ridge National Laboratory, Oak
  Ridge, TN 37831}
\author{M. Liebend\"orfer}
\affiliation{Canadian Institute for Theoretical Astrophysics, Toronto
  ON M5S 3H8}
\affiliation{Physics Division, Oak Ridge National Laboratory, Oak
  Ridge, TN 37831}
\affiliation{Department of Physics and Astronomy, University of
  Tennessee, Knoxville TN 37996}
\author{H.-Th. Janka}
\affiliation{Max-Planck-Institut f\"ur Astrophysik, D-85741 Garching,
  Germany}
\author{M. Rampp}
\affiliation{Max-Planck-Institut f\"ur Astrophysik, D-85741 Garching,
  Germany}

\date{\today}

\begin{abstract}
  Supernova simulations to date have assumed that during core collapse
  electron captures occur dominantly on free protons, while captures
  on heavy nuclei are Pauli-blocked and are ignored.  We have
  calculated rates for electron capture on nuclei with mass numbers
  $A=65$--112 for the temperatures and densities appropriate for core
  collapse. We find that these rates are large enough so that, in
  contrast to previous assumptions, electron capture on nuclei
  dominates over capture on free protons. This leads to significant
  changes in core collapse simulations.
\end{abstract}

\maketitle

At the end of their lives, stars with masses exceeding roughly
10~M$_\odot$ reach a moment in their evolution when their iron core
provides no further source of nuclear energy generation. At this time,
they collapse and, if not too massive, bounce and explode in
spectacular events known as type II or Ib/c supernovae. As the
density, $\rho$, of the star's center increases, electrons become more
degenerate and their chemical potential $\mu_e$ grows ($\mu_e \sim
\rho^{1/3}$). For sufficiently high values of the chemical potential
electrons are captured by nuclei producing neutrinos, which for
densities $\lesssim 10^{11}$~g~cm$^{-3}$, freely escape from the star,
removing energy and entropy from the core. Thus the entropy stays low
during collapse ensuring that nuclei dominate in the composition over
free protons and neutrons. During the presupernova stage, i.e.\ for
core densities $\lesssim 10^{10}$~g~cm$^{-3}$ and proton-to-nucleon
ratios $Y_e \gtrsim 0.42$, nuclei with $A = 55$--65 dominate. The
relevant rates for weak-interaction processes (including $\beta^{\pm}$
decay and electron and positron capture) were first estimated by
Fuller, Fowler and Newman~\cite{FFN} (for nuclei with $A<60$),
considering that at such conditions allowed (Fermi and Gamow-Teller)
transitions dominate. The rates have been recently improved based on
modern data and state-of-the-art many-body models~\cite{LMP},
considering nuclei with $A=45$--65. (This rate set will be denoted LMP
in the following.)  Presupernova models utilizing these improved weak
rates are presented in~\cite{Heger01}. In collapse simulations, i.e.\
densities $\gtrsim 10^{10}$~g~cm$^{-3}$, a much simpler description
of electron capture on nuclei is used. Here the rates are estimated
in the spirit of the independent particle model (IPM), assuming pure
Gamow-Teller (GT) transitions and considering only single particle
states for proton and neutron numbers between
$Z,N=20$--40~\cite{Bruenn85}. In particular this model results in
vanishing electron capture rates on nuclei with neutron numbers larger
than $N=40$, motivated by the observation~\cite{Fuller82} that, within
the IPM, GT transitions are Pauli-blocked for nuclei with $N \ge 40$
and $Z\le40$.

During core collapse, temperatures and densities are high enough to
ensure that nuclear statistical equilibrium (NSE) is achieved. This
means that for sufficiently low entropies, the matter composition is
dominated by the nuclei with the highest binding energy for a given
$Y_e$.  Electron capture reduces $Y_e$, driving the nuclear
composition to more neutron rich and heavier nuclei, including those
with $N>40$, which dominate the matter composition for densities
larger than a few $10^{10}$~g~cm$^{-3}$.  As a consequence of the
model applied in previous collapse simulations, electron capture
on nuclei ceases at these densities and the capture is entirely due to
free protons. We will show now that the employed model for electron
capture on nuclei is incorrect, as the Pauli-blocking of the GT
transitions is overcome by correlations~\cite{Langanke01} and
temperature effects~\cite{Fuller82,Wambach84}.

The residual nuclear interaction, beyond the IPM, mixes the $pf$ shell
with the levels of the $sdg$ shell, in particular with the lowest
orbital, $g_{9/2}$. This makes the closed $g_{9/2}$ orbit a magic
number in stable nuclei ($N=50$) and introduces, for example, a very
strong deformation in the $N=Z=40$ nucleus $^{80}$Zr. Moreover, the
description of the B(E2,$0^+ \rightarrow 2_1^+$) transition in
$^{68}$Ni requires configurations where more than one neutron is
promoted from the $pf$ shell into the $g_{9/2}$ orbit~\cite{Sorlin02},
unblocking the GT transition even in this proton-magic $N=40$ nucleus.
Such a non-vanishing GT strength has already been observed for
$^{72}$Ge ($N=40$)~\cite{Triumf} and $^{76}$Se ($N=42$)~\cite{Helmer}.
In addition, during core collapse electron capture on the nuclei of
interest occurs at temperatures $T\gtrsim0.8$~MeV, which, in the
Fermi gas model, corresponds to a nuclear excitation energy $U \approx
A T^2/8 \gtrsim 5$~MeV; this energy is noticeably larger than the
splitting of the $pf$ and $sdg$ orbitals ($E_{g_{9/2}}
-E_{p_{1/2},f_{5/2}} \approx 3$~MeV).  Hence, the configuration mixing
of $sdg$ and $pf$ orbitals will be rather strong in those excited
nuclear states of relevance for stellar electron capture.
Furthermore, the nuclear state density at $E\sim5$ MeV is already
larger than 100/MeV, making a state-by-state calculation of the rates
impossible, but also emphasizing the need for a nuclear model which
describes the correlation energy scale at the relevant temperatures
appropriately.  This model is the Shell Model Monte Carlo (SMMC)
approach~\cite{SMMC} which allows the calculations of nuclear
properties at finite temperature in unprecedentedly large model spaces.
To calculate electron capture rates for nuclei $A=65$--112 we have
first performed SMMC calculations in the full $pf$-$sdg$ shell, using
a residual pairing+quadrupole interaction, which, in this model space,
reproduces well the collectivity around the $N=Z=40$ region and the
observed low-lying spectra in nuclei like $^{64}$Ni and $^{64}$Ge.
From the SMMC calculations we determined the temperature-dependent
occupation numbers of the various single-particle orbitals, which then
became the input in RPA calculations of the capture rate, where we
considered allowed and forbidden transitions up to multipoles $J=4$,
including the momentum dependence of the operators.  This model is
described in more details in~\cite{Langanke01}, where, however, a
smaller model space has been used.


To validate our method at the early collapse conditions, we have
performed diagonalization shell model studies for $^{64,66}$Ni,
considering the complete $(pf)$ shell for $^{64}$Ni, and adopting, for
$^{66}$Ni, the $(pf)$ shell for protons and the $(pf_{5/2}g_{9/2})$
shell for neutrons. We find agreement to better than a factor 2
between the present (SMMC+RPA) rates and the diagonalization shell
model rates at stellar conditions ($T \lesssim 0.8$~MeV), for which
the latter can still be evaluated.

\begin{figure}[htb]
  \begin{center}
    \leavevmode
    \includegraphics[width=\columnwidth]{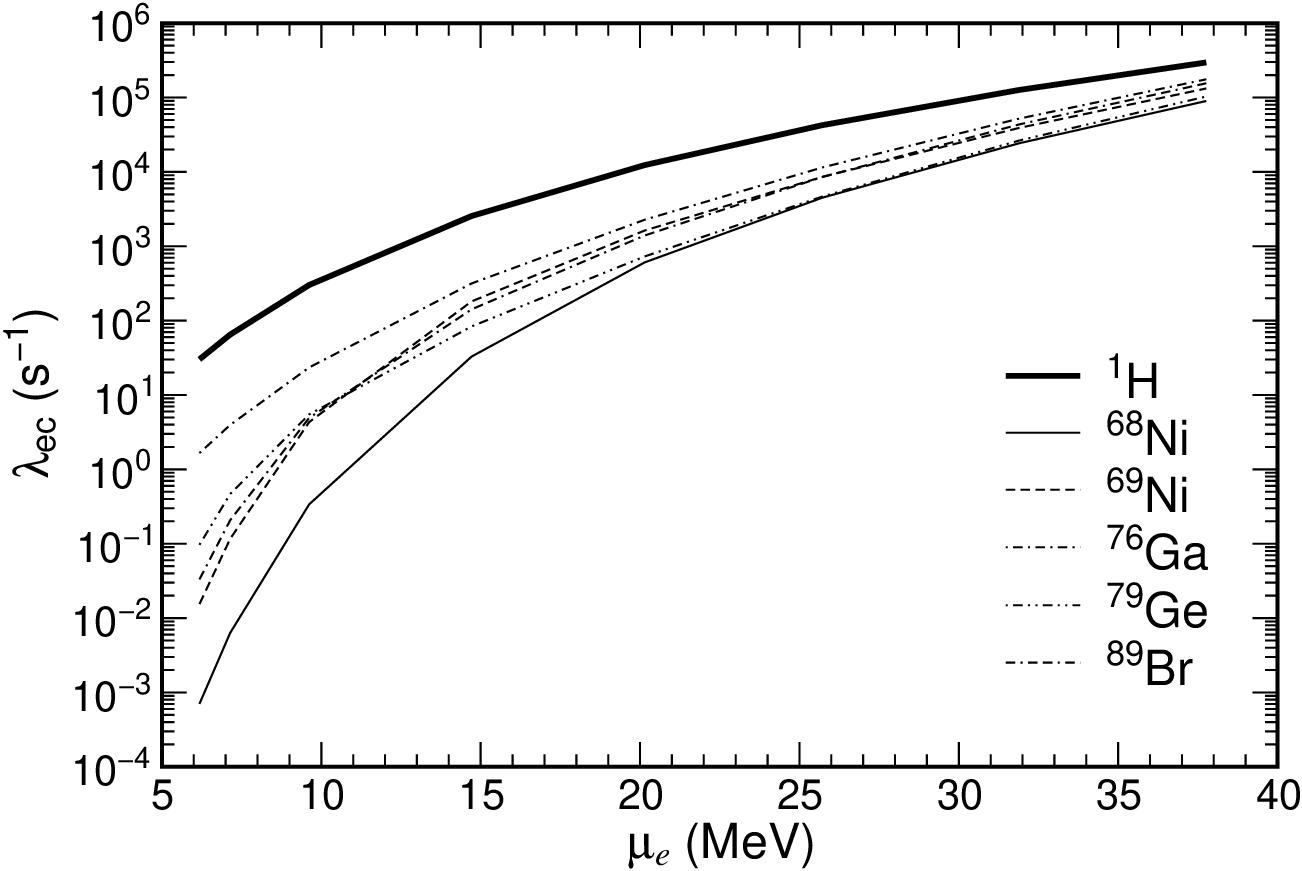}
    \caption{Comparison of the electron capture rates
      on free protons and selected nuclei as function of the electron
      chemical potential along a stellar collapse trajectory taken
      from~\cite{Mezzacappa01}. Neutrino blocking of the phase space
      is not included in the calculation of the
      rates.\label{fig:individual}}
  \end{center}
\end{figure}

\begin{figure}[htb]
  \begin{center}
    \leavevmode
    \includegraphics[width=1.0\columnwidth]{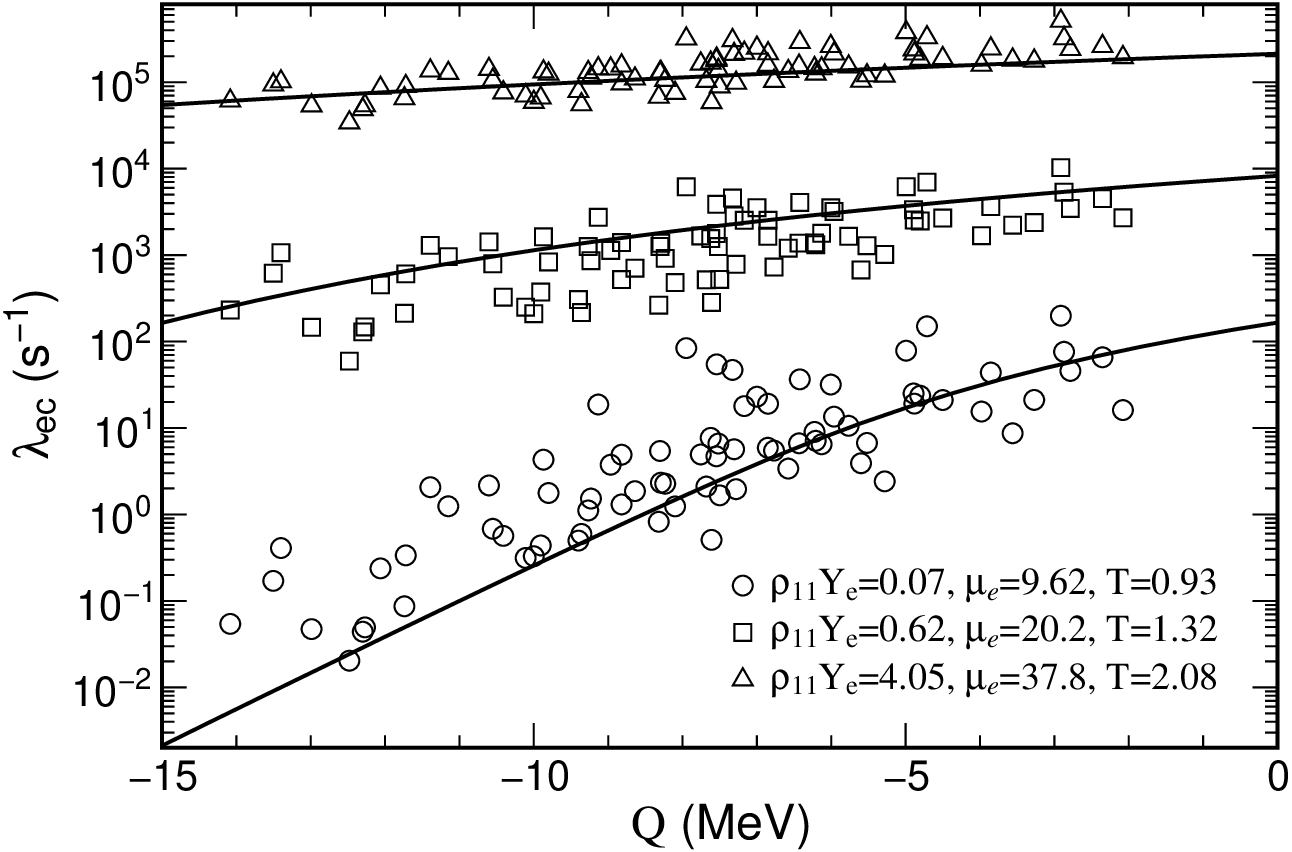}
    \caption{Electron capture rates on nuclei as function of $Q$-value
      for 3 different stellar conditions. Temperature and electron
      chemical potential are measured in MeV.  The solid lines
      represent the approximate $Q$-dependence of the rates as defined
      in Eq.~(\ref{eq:lambda}). Neutrino blocking of the phase space
      is not included in the calculation of the rates. $\rho_{11}$
      measures the density in units of $10^{11}$~g~cm$^{-3}$.
      \label{fig:qvalue}}
      \end{center}
\end{figure}

For all studied nuclei we find neutron holes in the $(pf)$ shell and,
for $Z >30$, non-negligible proton occupation numbers for the $sdg$
orbitals. This unblocks the GT transitions and leads to sizable
electron capture rates. Fig.~\ref{fig:individual} compares the
electron capture rates for free protons and selected nuclei along a
core collapse trajectory, as taken from~\cite{Mezzacappa01}. Depending
on their proton-to-nucleon ratio $Y_e$ and their $Q$-values, these
nuclei are abundant at different stages of the collapse. For all
nuclei, the rates are dominated by GT transitions at low densities,
while forbidden transitions contribute sizably for $\gtrsim
10^{11}$~g~cm$^{-3}$.  The electron chemical potential $\mu_e$ and the
reaction $Q$-value are the two important energy scales of the capture
process.  At a given density, i.e.\ constant $\mu_e$, the rate is
generally larger for nuclei with smaller $Q$-values.  The rate is
sensitive to the GT strength distribution, if $\mu_e \lesssim Q$.
However, $\mu_e$ increases much faster with density than the
$Q$-values of the abundant nuclei. As a consequence the capture rates
on nuclei become quite similar at larger densities, say $\gtrsim
10^{11}$~g~cm$^{-3}$, depending now basically only on the total GT
strength, but not its detailed distribution.  This is demonstrated in
Fig.~\ref{fig:qvalue} which shows our calculated capture rates as
function of $Q$-value at 3 different stellar conditions.  The
$Q$-value dependence of the capture rate for a transition from a
parent state at excitation energy $E_i$ to a daughter state at $E_f$
($\Delta E = E_f-E_i)$ is well approximated by \cite{FFN85}
\begin{equation}
\label{eq:lambda}
\lambda = \frac{(\ln 2)B}{K} \left(\frac{T}{m_e c^2}\right)^5
\left[F_4(\eta) - 2  
\chi F_3(\eta) +
\chi^2 F_2 (\eta)\right]
\end{equation}
where $\chi = (Q-\Delta E)/T$, $\eta=(\mu_e + Q - \Delta E)/T$,
$K=6146$~s and $B$ represents a typical (Gamow-Teller plus forbidden)
matrix element. The quantities $F_k$ are the relativistic Fermi
integrals of order $k$.
  
At ($\rho Y_e=7\times10^9\text{ g cm}^{-3}, T=0.93$ MeV), we observe
some scatter of the calculated rates around the mean $Q$-dependence
indicating that several parent and daughter states with different
transition strengths contribute. For nuclei with large $|Q|$-values
the simple parametrization breaks down. However, at $\rho_{11} Y_e
\gtrsim 4$, the electron chemical potential has increased sufficiently
that the rates become virtually independent of the strength
distribution and are well represented by the average $Q$-value
dependence~(\ref{eq:lambda}) with $B=4.6$ and $\Delta E =2.5$ MeV.
Such a parametrization could then be adopted in core collapse
simulations for even higher densities, when nuclei heavier than the
ones included in the present study start to dominate in the
composition.

\begin{figure}[htb]
  \begin{center}
    \leavevmode
    \includegraphics[width=1.0\columnwidth]{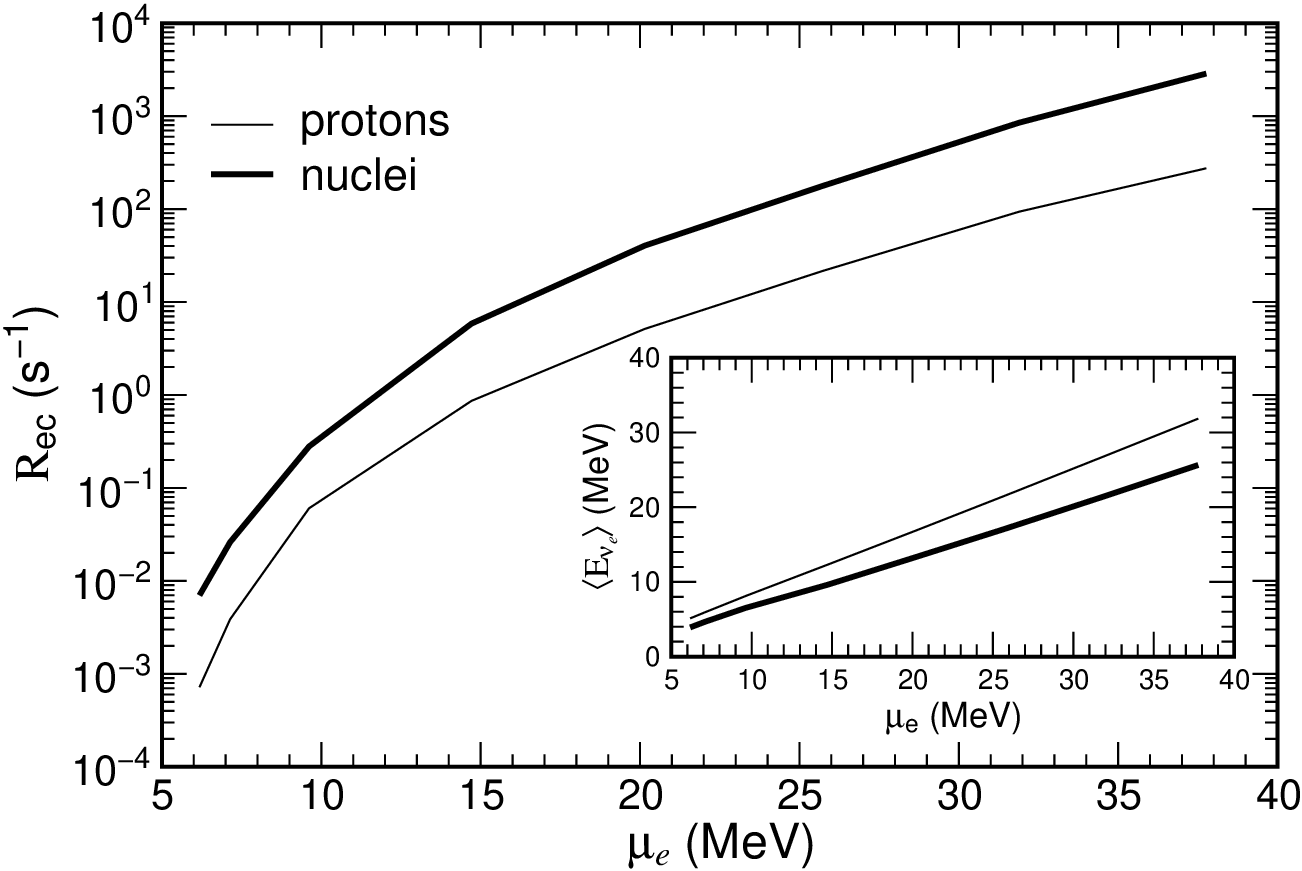}
    \caption{The reaction rates for electron capture on protons (thin
      line) and nuclei (thick line) are compared as a function of
      electron chemical potential along a stellar collapse trajectory
      taken from~\cite{Mezzacappa01}.  The insert shows the related
      average energy of the neutrinos emitted by capture on nuclei and
      protons. The results for nuclei are averaged over the full
      nuclear composition (see text).  Neutrino blocking of the phase
      space is not included in the calculation of the rates.
      \label{fig:average}}
  \end{center}
\end{figure}

Simulations of core collapse require reaction rates for electron
capture on protons, $R_p= Y_p \lambda_p$, and nuclei $R_h = \sum_i Y_i
\lambda_i$ (where the sum runs over all the nuclei present and $Y_i$
denotes the number abundance of a given species), over wide ranges in
density and temperature.  While $R_p$ is readily derived from
\cite{Bruenn85}, the calculation of $R_h$ requires knowledge of the
nuclear composition, in addition to the electron capture rates
described earlier. The information about the nuclear composition
provided by the commonly used Lattimer-Swesty equation of
state~\cite{LaSw91}, the total abundance of heavy nuclei and the
average $Z$ and $A$, is not sufficiently detailed to make adequate use
of these new reaction rates.  Therefore a Saha-like NSE is used to
calculate the needed abundances of individual isotopes, including
Coulomb corrections to the nuclear binding energy
\cite{HixPhD,BrGa99}, but neglecting the effects of degenerate
nucleons \cite{ElHi80}.  The combination of this NSE with electron
capture rates for approximately 200 nuclei with $A=45$--112, which we
have determined here and in Ref.~\cite{LMP}, was used to compute the
rate of electron capture on nuclei and the emitted neutrino spectra as
a function of temperature, density and electron fraction. This is
similar to treatments used in investigations of electron capture
during stellar evolution~\cite{Heger01} and in thermonuclear
supernovae~\cite{Brachwitz00}. The rates for the inverse
neutrino-absorption process are determined from the electron capture
rates by detailed balance. Due to its much smaller $|Q|$-value, the
electron capture rate on the free protons is larger than the rates of
abundant nuclei during the core collapse (fig.~\ref{fig:individual}).
However, this is misleading as the low entropy keeps the protons
significantly less abundant than heavy nuclei during the collapse.
Fig.~\ref{fig:average} shows that the reaction rate on nuclei, $R_h$,
dominates the one on protons, $R_p$, by roughly an order of magnitude
throughout the collapse when the composition is considered. Only after
the bounce shock has formed does $R_p$ become higher than $R_h$, due
to the high entropies and high temperatures in the shock-heated matter
that result in a high proton abundance. The obvious conclusion is that
electron capture on nuclei must be included in collapse simulations.



It is also important to stress that electron capture on nuclei and on
free protons differ quite noticeably in the neutrino spectra they
generate. The average neutrino energy, $\langle E_\nu \rangle$, of the
neutrinos emitted by electron capture on nuclei, can be obtained
dividing the neutrino energy loss rate (defined by $\sum_i Y_i
\mathcal{E}_i$ where $\mathcal{E}_i$ is the energy loss rate by
electron capture on nucleus $i$ and $Y_i$ denotes its number
abundance) by the reaction rate for electron capture on nuclei, $R_h$.
For the neutrino spectrum we adopt the parametrized form as defined
in~\cite{Langanke01b}, adjusted to reproduce the average neutrino
energy $\langle E_\nu\rangle$. The neutrino emissivity is then
obtained by multiplying the NSE-averaged electron capture rate by the
neutrino spectra.  

Fig.~\ref{fig:average} demonstrates that neutrinos from captures on
nuclei have a mean energy 40--60\% less than those produced by capture
on protons.  Although capture on nuclei under stellar conditions
involves excited states in the parent and daughter nuclei, it is
mainly the larger $|Q|$-value which significantly shifts the energies
of the emitted neutrinos to smaller values. Despite that, the total
neutrino energy loss rate is larger when electron capture on nuclei is
considered, caused by the increase in the total (nuclei plus protons)
electron capture rate.  The differences in the neutrino spectra
strongly influence neutrino-matter interactions, which scale with the
square of the neutrino energy and are essential for collapse
simulations. In current simulations~\cite{Mezzacappa01,Janka01}, the
low energy portions of the neutrino distribution are populated via
neutrino-electron inelastic scattering of high energy neutrinos
produced by electron capture on free protons. Electron capture on
nuclei produces neutrinos with significantly lower energies,
accelerating this redistribution process. In this context, inelastic
neutrino-nucleus scattering, which is usually ignored, could also be
an important process~\cite{haxton91,sampaio02}.

\begin{figure}[htb]
  \includegraphics[width=1.0\columnwidth]{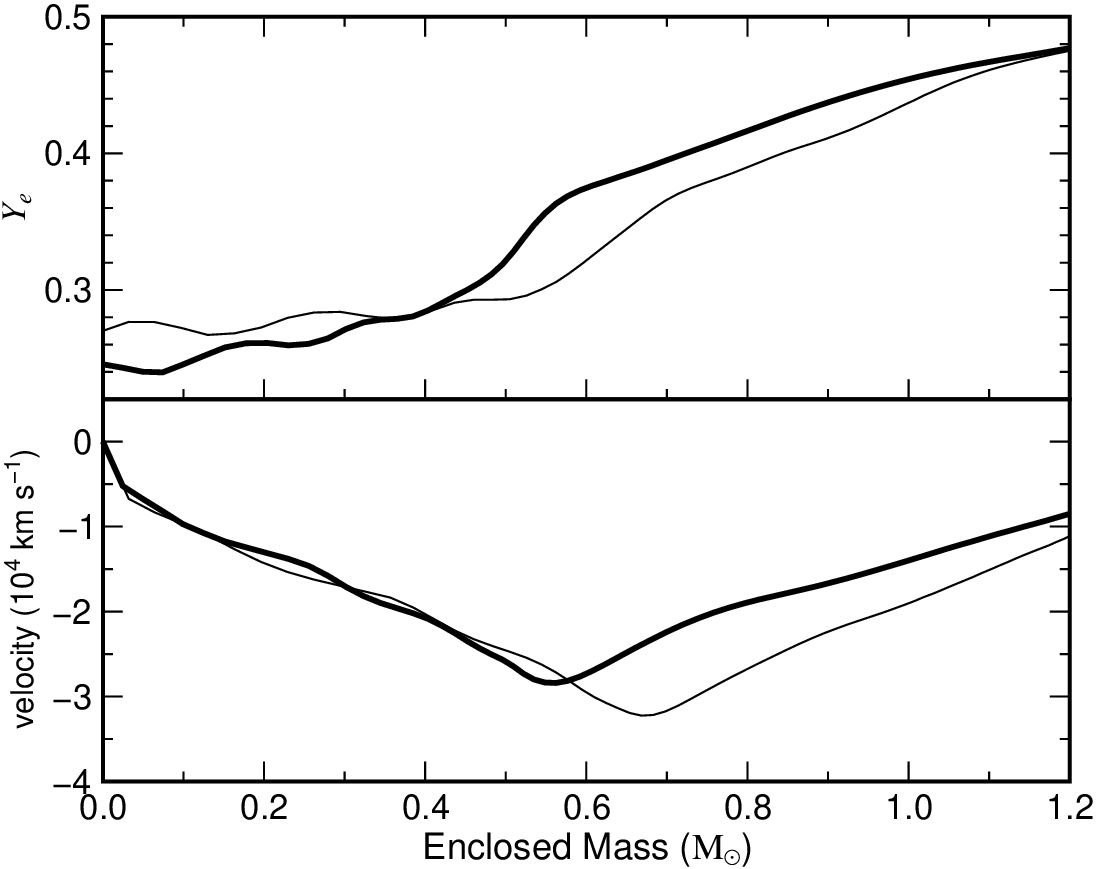}
  \caption{The electron fraction and velocity as functions of the 
    enclosed mass at the moment when the center reaches nuclear matter
    densities for a 15~M$_\odot$ model \cite{Heger01}.  The thin line
    is a simulation using the Bruenn parameterization~\cite{Bruenn85}
    while the thick line is for a simulation using the combined
    LMP~\cite{LMP} and SMMC+RPA rate sets. Both models were calculated
    with Newtonian gravity. \label{fig:bounce}}
\end{figure}

The effects of this more realistic implementation of electron capture
on heavy nuclei have been evaluated in independent self-consistent
neutrino radiation hydrodynamics simulations by the Oak Ridge and
Garching collaborations \cite{Hix03,Janka03}. The basis of these
models is described in detail in Refs.~\cite{Mezzacappa01} and
\cite{Janka01}.  Both collapse simulations yield qualitatively the
same results.  Here we show a key result obtained by the Oak Ridge
collaboration demonstrating that the effects of this improved
treatment of nuclear electron capture are twofold. In regions close to
the center of the star, the additional electron capture on heavy
nuclei results in more electron capture in the new models.  In regions
where nuclei with $A<65$ dominate, the LMP rates result in less
electron capture.  The results of these competing effects can be seen
in the first panel of Figure~\ref{fig:bounce}, which shows the
distribution of $Y_e$ throughout the core when the central density
reaches $10^{14}$~g~cm$^{-3}$, making the transition to nuclear
matter.  The combination of increased electron capture in the interior
with reduced electron capture in the outer regions displaces the
velocity minimum, which marks the eventual location of shock
formation, by 0.1~M$_\odot$. The full effects of these changes on the
bounce and post-bounce evolution in supernova models will be discussed
in~\cite{Hix03,Janka03}.

Our calculations clearly show that the many neutron-rich nuclei which
dominate the nuclear composition throughout the collapse of a massive
star also dominate the rate of electron capture.  Astrophysics
simulations have demonstrated that these rates have a strong impact on
the core collapse trajectory and the properties of the core at bounce.
The evaluation of the rates has to rely on theory as a direct
experimental determination of the rates for the relevant stellar
conditions (i.e.\ rather high temperatures) is currently impossible.
Nevertheless it is important to experimentally explore the
configuration mixing between $pf$ and $sdg$ shell in extremely
neutron-rich nuclei as such understanding will guide and severely
constrain nuclear models. Such guidance is expected from future
radioactive ion-beam facilities.

\begin{acknowledgments}
  The work has been partly supported by the Danish Research Council,
  by the Spanish MCYT under contract AYA2002-04094-C03-02, by NASA
  under contract NAG5-8405, by the National Science Foundation under
  contract AST-9877130, by the Department of Energy, through the
  PECASE and Scientic Discovery through Advanced Computing Programs
  and by funds from the Joint Institute for Heavy Ion Research.  Oak
  Ridge National Laboratory is managed by UT-Battelle, LLC, for the
  U.S. Department of Energy under contract DE-AC05-00OR22725.  JMS
  acknowledges the financial support of the ``Funda\c{c}\~ao para a
  Ci\^encia e Tecnologia''. HTJ and MR acknowledge support by the
  Sonderforschungsbereich 375 ``Astro-Teilchenphysik'' of the Deutsche
  Forschungsgemeinschaft. KL and HTJ thank the ECT* in Trento, where
  part of this collabotation has been initiated, for its hospitality.
\end{acknowledgments}

\end{document}